\documentclass[aps, prx, twocolumn,10pt,showpacs,superscriptaddress,amsmath,amssymb,nobibnotes]{revtex4-2}

\usepackage{graphicx}
\usepackage{amsfonts}    
\usepackage{verbatim}   
\usepackage{color}      
\usepackage{subfigure}  
\usepackage{hyperref}   
\usepackage{booktabs}
\usepackage{threeparttable}
\usepackage{dcolumn}
\usepackage{multirow}
\usepackage{physics}
\usepackage[most]{tcolorbox}
\usepackage[capitalize]{cleveref}

\tcbuselibrary{theorems}

\newtcbtheorem[number within=section]{mybox}{Box}{
    halign title=left, 
    halign=left, 
}{bx}

\usepackage{enumitem}


\usepackage{graphicx}
\graphicspath{{figures/}}
\usepackage{booktabs}

\usepackage{cleveref}

\begin{document}

\title{ Fault-Tolerant Quantum Key Distribution: Enabling Overclocked Modulation  }

\author{Feng-Yu Lu}\email{These authors contributed equally to this work}
\author{Jia-Xuan Li}\email{These authors contributed equally to this work}
\author{Ze-Hao Wang}\email{These authors contributed equally to this work}
\affiliation{CAS Key Laboratory of Quantum Information, University of Science and Technology of China, Hefei, Anhui 230026, P. R. China}
\affiliation{CAS Center for Excellence in Quantum Information and Quantum Physics, University of Science and Technology of China, Hefei, Anhui 230026, P. R. China}
\author{Shuang Wang}\email{wshuang@ustc.edu.cn}
\author{Zhen-Qiang Yin}\email{yinzq@ustc.edu.cn}
\affiliation{CAS Key Laboratory of Quantum Information, University of Science and Technology of China, Hefei, Anhui 230026, P. R. China}
\affiliation{CAS Center for Excellence in Quantum Information and Quantum Physics, University of Science and Technology of China, Hefei, Anhui 230026, P. R. China}
\author{Álvaro Navarrete}
\author{Marcos Curty}
\affiliation{Vigo Quantum Communication Center, University of Vigo, Vigo E-36310, Spain}
\affiliation{Escuela de Ingeniería de Telecomunicación, Department of Signal Theory and Communications, University of Vigo, Vigo E-36310, Spain}
\affiliation{AtlanTTic Research Center, University of Vigo, Vigo E-36310, Spain}
\author{Wei Chen}
\author{De-Yong He}
\author{Guang-Can Guo}
\author{Zheng-Fu Han}
\affiliation{CAS Key Laboratory of Quantum Information, University of Science and Technology of China, Hefei, Anhui 230026, P. R. China}
\affiliation{CAS Center for Excellence in Quantum Information and Quantum Physics, University of Science and Technology of China, Hefei, Anhui 230026, P. R. China}
\affiliation{Hefei National Laboratory, University of Science and Technology of China, Hefei 230088, China}
\date{\today}

\begin{abstract}
Implementation security, higher generation rate, and lower cost are primary missions in the domain of quantum key distributions in recent years. However, simultaneously achieving robust security, high speed, and low cost often resembles an ``impossible triangle''. This is largely because the modulation system, a core component of the QKD transmitter, imposes a strict bandwidth limitation. Pushing a low-cost modulator to a high repetition frequency inevitably introduces correlations and misalignment, which can create security loopholes. Conversely, operating at a conservative rate fails to exploit the system's potential, while adopting ultra-high-bandwidth components is often expensive for practical implementation, forcing a perpetual trade-off among implementation security, key rate, and cost. In this work, we propose a comprehensive countermeasure to overcome this modulation bandwidth bottleneck. We present a protocol specifically designed to address the security loopholes arising from modulation imperfections, ensuring security even in overclocked modulation systems. Furthermore, we develop two practical techniques to characterize and mitigate the detrimental correlations. Our experimental setup demonstrates that the proposed method achieves the lowest correlated deviation reported in similar studies, while maintaining a high secret key rate using a bandwidth-limited modulation system. By simultaneously enhancing security, performance, and practicality, this work releases QKD systems from the traditional performance-cost trade-off in the near term, paving the way for widespread deployment. In the long run, this work can be readily integrated with high-bandwidth components to further push the boundaries of system performance.
\end{abstract}


\maketitle

\section{\bf INTRODUCTION} 

As one of the most successful technologies in the field of quantum information science, quantum key distribution (QKD) \cite{BB84,lo1999unconditional,shor2000simple,scarani2009security,Rennersecurity} allows two distant parties, usually referred to as Alice and Bob, to share information-theoretically secure keys in the presence of a technologically unbounded eavesdropper, Eve.

As a technology for encryption, security is the fundamental requirement for any QKD setup \cite{lo2012measurement,braunstein2012side,tamaki2014loss,xu2015discrete,nagamatsu2016security,wang2019practical,lu2022unbalanced,tomamichel2012tight,sajeed2016insecurity,curty2014finite,pirandola2015high,tamaki2016decoy,wang2023fully,george2021numerical,wang2022numerical,navarrete2021practical};
as a system for communication, the pursuit of higher secret-key rates \cite{tfqkd,tanaka2012high,yuan201810,bacco2023high,li2023high,boaron2018secure,wang2022twin,liu2023experimental,grunenfelder2020performance} is the inevitable tendency for the development of the QKD domain; as a technique for practical applications, lower cost would be a great advantage for its deployment \cite{sax2023high,sibson2017chip,sibson2017integrated,wei2020high,zhang2022polarization}.
In practice, however, achieving high speed, robust security, and low cost simultaneously often resembles an `impossible triangle'. Many experiments \cite{yoshino2018quantum,roberts2018patterning,lu2021intensity,kang2022patterning,lu2023intensity,gao2023suppression,xing2024characterization,trefilov2025intensity} have shown that when low-cost modulation systems are used to force high-speed modulation, the QKD transmitter inevitably suffers from correlation and misalignment issues, which may create security loopholes if not properly addressed. Meanwhile, operating the system at a conservative frequency fails to fully exploit its potential, while high-bandwidth systems, although capable of high speed and security, remain expensive for practical deployment. This limitation forces us to navigate trade-offs among security, key rate, and cost.

\begin{figure*}
    \begin{mybox}{protocol procedure}{protocol}\label{box1}
        \begin{enumerate}[label=\arabic*., leftmargin=*]
            \item 
            \textbf{System calibration:}
            Before the quantum communication, Alice pre-decides her decoy-state intensity \cite{hwang2003quantum,wang2005beating,lo2005decoy} $\mathbb{A} = \{\mu, \nu, \omega\}$ and bit/basis $\mathbb{R} = \{ {0}, {1}, +\}$ settings. Then, she pre-measures the correlation range $\xi$, and characterizes the actual intensity $\alpha_{s_{k-\xi}^k}$ and the actual encoding $\iota_{s_{k-\xi}^k}$ of the emitted phase-randomized weak coherent pulses for each possible setting pattern $s_{k-\xi}^k\in \mathbb{A}^{\xi}\times\mathbb{R}^{\xi}$ of length $\xi$. That is, the pattern $s_{k-\xi}^{k} \equiv s_{k}s_{k-1}...s_{k-\xi}$, with $s_k \equiv (a_k,r_k)$, includes not only Alice's intensity setting $a_k\in \mathbb{A}$ and bit/basis setting $r_k\in \mathbb{R}$ associated with the round $k$ in which the pulse is emitted, but also the settings associated with the previous $\xi$ rounds.
            \item \textbf{Quantum Communication:} In each round $k\in\{1,\dots,N\}$ of the protocol, the parties do the following:
            
            \begin{enumerate}[label=\roman*., leftmargin=*]
                \item 
                \textbf{State preparation:}
                Alice selects an intensity (bit/basis) setting $a_k \in \mathbb{A}$ ($r_k \in \mathbb{R}$) with probability $p_{a_k}$ ($p_{r_k}$), and \textit{tries} to prepare a phase-randomized weak coherent pulse (PRWCP) accordingly.
                The $Z=\{0,1\}$ basis, which she selects with probability $P_Z^A=p_0+p_1$, is used for key generation, while the $X=\{+\}$ basis, which she selects with probability $P_X^A=p_+$, is used for testing the channel.  
                \item 
                \textbf{Measurement:}
                Bob randomly selects a measurement basis $x_k \in \{Z, X\}$ with probability $P_Z^B$ and $P_X^B$ to measure the incoming signal, and he records the measurement outcome $\kappa_k \in \{\varnothing,0,1\}$, where $\varnothing$ represents the no-detection event.
            \end{enumerate}  
            \item 
            \textbf{Sifting:}
            Alice and Bob broadcast their basis selection for each round, and Bob further announces if the round was detected or not. 
            Then they construct their sifted keys from a random subset of their bits $r_k$ and $\kappa_k$ associated with the detected $Z$-basis rounds in which Alice selected the signal intensity $\mu$. All the remaining records are publicly revealed for parameter estimation.
            \item 
            \textbf{Parameter estimation:}
            Alice and Bob calculate the conditional gains $Q^{x,\kappa}_{s_{k-\xi}^k}$ and quantum bit error rates (QBERs) $E_{s_{k-\xi}^k}$, and employ our `enhanced decoy-state method' to lower bound the conditional single-photon yields $Y^{x,\kappa}_{s_{k-\xi}^k}$ and upper bound the single-photon error rates $e_{s_{k-\xi}^k}$. 
            \item 
            \textbf{Key distillation:}
            Alice and Bob perform error correction, error verification and privacy amplification to generate two identical secret keys.
        \end{enumerate}
    \end{mybox}
\end{figure*}

Numerous efforts have been undertaken to address this challenge. Some efforts \cite{laing2010reference,tamaki2014loss,hwang2017improved,mizutani2019quantum,lu2022unbalanced} efficiently mitigate the impact of static state preparation flaws (SPFs). Other efforts address side channels resulting from mode dependencies (including Trojan-horse attacks) \cite{wang2019practical,pereira2019quantum}. Furthermore, several protocols or proofs tackle correlated modulation, also known as the patterning effect \cite{yoshino2018quantum,pereira2020quantum,zapatero2021security,sixto2022security,kang2022patterning,lu2023intensity}. 
However, these studies typically address only one or a few of the aforementioned issues in isolation: most works fail to account for the presence of correlations; some studies only consider correlated intensity sources \cite{yoshino2018quantum,zapatero2021security,sixto2022security}; and some focus solely on correlated bit and basis encoding in ideal single-photon systems \cite{pereira2020quantum}. Moreover, some studies have attempted to address this problem from a technical standpoint, including data post-processing \cite{yoshino2018quantum}, pre-processing \cite{kang2022patterning,lu2023intensity}, and correlation-mitigating modulators \cite{roberts2018patterning,lu2021intensity,gao2023suppression}. However, these latter approaches only consider intensity correlations and their performance requires further improvement. 

To address the challenge of the `impossible triangle', it is essential to account for all the above imperfections together to realize a modulation overclocking. Additionally, characterizing, measuring, and subsequently suppressing these imperfections is crucial for avoiding the performance decrease. 
In this study, we introduce a fault-tolerant protocol that can handle SPFs, mode-dependent side channels, and pulse correlations holistically, thereby allowing QKD systems to work at a higher frequency without losing their security. Furthermore, our protocol outperforms previous approaches with bandwidth-limited modulation systems. In addition, we have developed several techniques in this study to reduce the misalignment and correlation errors, thus avoiding the decrease of SKR. One of the techniques, named `deviation microscope', successfully addresses the challenge of measuring weak-intensity correlations. This enables the measurement of correlations of the vacuum state and time-bin encoding. Another technique, termed `double suppressing', mitigates correlated deviations to an ultra-low level, which represents the state-of-the-art suppression when compared to other similar works \cite{yoshino2018quantum,roberts2018patterning,lu2021intensity,kang2022patterning,lu2023intensity,gao2023suppression}. Importantly, our theory provides a precise estimation of information leakage, while our techniques minimize misalignment and correlated errors in bandwidth-limited modulators. Therefore, both security and performance aspects are addressed through our theoretical and technical advancements. Building on these achievements, we experimentally demonstrate an overclocked QKD system and successfully overcome the key rate limitations imposed by modulation bandwidth. In summary, our work makes it possible to develop high-quality QKD systems while reducing their complexity and cost, thus paving the avenue for QKD's practical applications, especially for further field and network applications. Meanwhile, in the long run, this work can be readily integrated with high-bandwidth components to further push the boundaries of system performance.

\section{\bf Theoretical framework} 
Based on previous theoretical and experimental results \cite{pereira2020quantum,zapatero2021security,sixto2022security,yoshino2018quantum,roberts2018patterning,lu2021intensity,kang2022patterning,lu2023intensity,gao2023suppression}, we introduce a practical transmitter model. This model accounts for all the aforementioned imperfections with a small set of assumptions, representing a significant advancement over previous models.

Our protocol is described as follows (see Box. 1): In each round $k$, Alice randomly selects a bit/basis encoding setting $r_k\in \mathbb{R}=\{0,1,+\}$ and an intensity setting $a_k\in \mathbb{A}=\{\mu,\nu,\omega\}$ and prepares a phase-randomized coherent state accordingly. However, due to the modulation bandwidth limitation, both the actual encoding $\iota_k$ and the actual intensity $\alpha_k$ of the transmitted pulse may differ from her ideal selection and depend on the settings selected in previous rounds.  In particular, here we shall consider that this dependence has a finite range $\xi$ (referred to as $\xi$-order correlation) meaning that $\iota_k$ and $\alpha_k$ may depend on the settings $r_j$ and $a_j$ with $j\in\{k,\dots,k-\xi\}$, but they are unaffected by those with $j<k-\xi$ \cite{pereira2020quantum,zapatero2021security,sixto2022security,kang2022patterning}. Importantly, we note that the security analysis could be extended to the case of an infinite correlation length by incorporating the results of~\cite{pereira2024quantum}. Most previous works typically assume the actual bit/basis preparation (intensity) is solely influenced by the previous bit/basis encoding (intensity) settings, expressed as $\iota_k \equiv \iota_{r_{k-\xi}^{k}}$ and $\alpha_k \equiv \alpha_{a_{k-\xi}^{k}}$. Phase-encoding and polarization-encoding schemes typically adhere to this scenario, as phase and polarization modulation are often regarded as independent of the intensity modulation. 
In this work, we consider the more general scenario in which both $\iota_k$ and $\alpha_k$ may be affected by the full sequence of previous settings $s_{k-\xi}^k$, expressed as $\iota_k \equiv \iota_{s_{k-\xi}^{k}}$ and $\alpha_k \equiv \alpha_{s_{k-\xi}^{k}}$. Time-bin encoding usually adheres to this scenario, given that time-bin bit/basis encoding fundamentally involves intensity modulation~\cite{wang2015phase,yin2016measurement,boaron2018secure}. Another example that may adhere to this scenario is chip-based QKD, independently of the encoding~\cite{li2018secure,Xing2025}.

Based on the model of the bandwidth-limited modulations, we propose a fault-tolerant QKD protocol that remains secure in the presence of the aforementioned imperfections and considerably reduces the required assumptions (see Box. 1). For this, prior to the protocol execution, Alice accurately characterizes the quantum states of the transmitted pulses to determine the correlation range $\xi$, as well as the actual bit/basis encoding $\iota_{s_{k-\xi}^k}$ and the actual intensities $\alpha_{s_{k-\xi}^k}$ for each sequence $s_{k-\xi}^k$. This enables Alice and Bob to post-process their raw keys in a fine-grained manner. That is, they not only classify the measurement statistics based on the single-round setting choices $s_k$, but also take into account Alice's $\xi$ previous setting choices $s_{k-\xi}^{k-1}$.

Specifically, pulse correlations are incorporated into the model through a series of constraints that restrict the deviations between the actual intensity and state preparation from an idealized scenario with no correlations. These constraints are validated by Alice during the transmitter characterization step prior to the protocol execution, and serve as inputs for the security proof. We refer the reader to the Supplementary Information for further details.

Additionally, SPFs are accommodated via the rejected-data analysis~\cite{tamaki2014loss}, which allows to tightly estimate the detection statistics of some virtual states that are required to compute the phase-error rate of the protocol. Importantly, due to the presence of information leakage, the states of the single-photon contributions do not lie in a qubit space, preventing us from directly calculating the phase-error rate.
To solve this, we rely on the so-called CS inequality~\cite{pereira2020quantum,zapatero2021security} to estimate the measurement statistics of an auxiliary state---that lies within the qubit space spanned by the $Z$-basis states---which is sufficiently close to the test state (\textit{i.e.}, to the single-photon state encoding $r_k=+$). Moreover, we employ a refined decoy-state method which uses linearized CS constraints~\cite{zapatero2021security,sixto2022security} to bound the single-photon yields and error rates in the presence of intensity correlations. 
By combining these tools, we can realize the modulation overclocking. 

In the asymptotic regime of infinitely many rounds, the SKR can be approximated as~\cite{zapatero2021security} 
\begin{equation}
    \label{equ:keyrate}
    \begin{aligned}
    K = p_{\mu} P_Z^{A} P_Z^{B} \left\{p_{1|\mu}^{\rm L}  y_Z^{\rm L}
         \left[1-h\left({e}_{\rm p}^{\rm U}\right)\right] - f Q_{\mu}^Z h\left({e}_{\rm b}\right) \right\},
   \end{aligned}
\end{equation}
where $p_{1|\mu}^L$, $y_Z^L$, and ${e}_{\rm p}^U$ denote, respectively, a lower bound on the probability of emitting a single-photon pulse when Alice selects the intensity setting $\mu$, a lower bound on the single-photon yield, and an upper bound on the single-photon phase-error rate (both in the $Z$ basis), averaged over all possible $\xi$-length setting sequences; $h(\cdot)$ denotes the Shannon binary entropy function; $f$ is the error correction efficiency; $Q_{\mu}^Z$ is the gain of the overall signal states in the $Z$-basis; and $e_{\rm b}$ is the quantum bit-error rate of the sifted key in the $Z$-basis.

To validate the performance of an overclocked system using our protocol, we simulate an ideal BB84 scheme with a maximum secure frequency of 250 MHz~\cite{kang2022patterning,lu2023intensity}---\textit{i.e.}, this is the highest frequency at which the system can operate without inducing pulse correlations---and compare it with a double-frequency overclocked system (500 MHz) and a quadruple-frequency overclocked system (1 GHz), introduce correlations of range 1 and 3, respectively. The systems with and without cross-correlation are both validated by simulation. The results indicate that the overclocked systems obtain significantly higher SKRs before approaching the maximum distance. In particular, Fig. 1a shows that the overclocked system has a SKR boost close to a multiple of the overclocking in the range of up to 10 dB (\textit{i.e.}, 50 km for standard fiber loss). Moreover, Fig. 1b shows that the quadruple-frequency overclocked system still has a 3-times higher SKR in the 10 dB range even if cross correlations are considered. 
In the Supplementary Information we define various different $\varepsilon$ parameters that characterize the strength of the correlations. For simplicity, in this figure we set all the different types of $\varepsilon$ to the same value $\epsilon$. Notably, for $\epsilon=10^{-6}$ the overclocked system still obtains a superior SKR at 25 dB total loss, which for standard optical fiber corresponds to $\sim50$ km and $\sim100$ km when using single-photon avalanche detectors (SPAD) and superconducting-nanowire single-photon detectors (SNSPD), respectively.

\begin{figure}
   \includegraphics[width=0.5\textwidth]{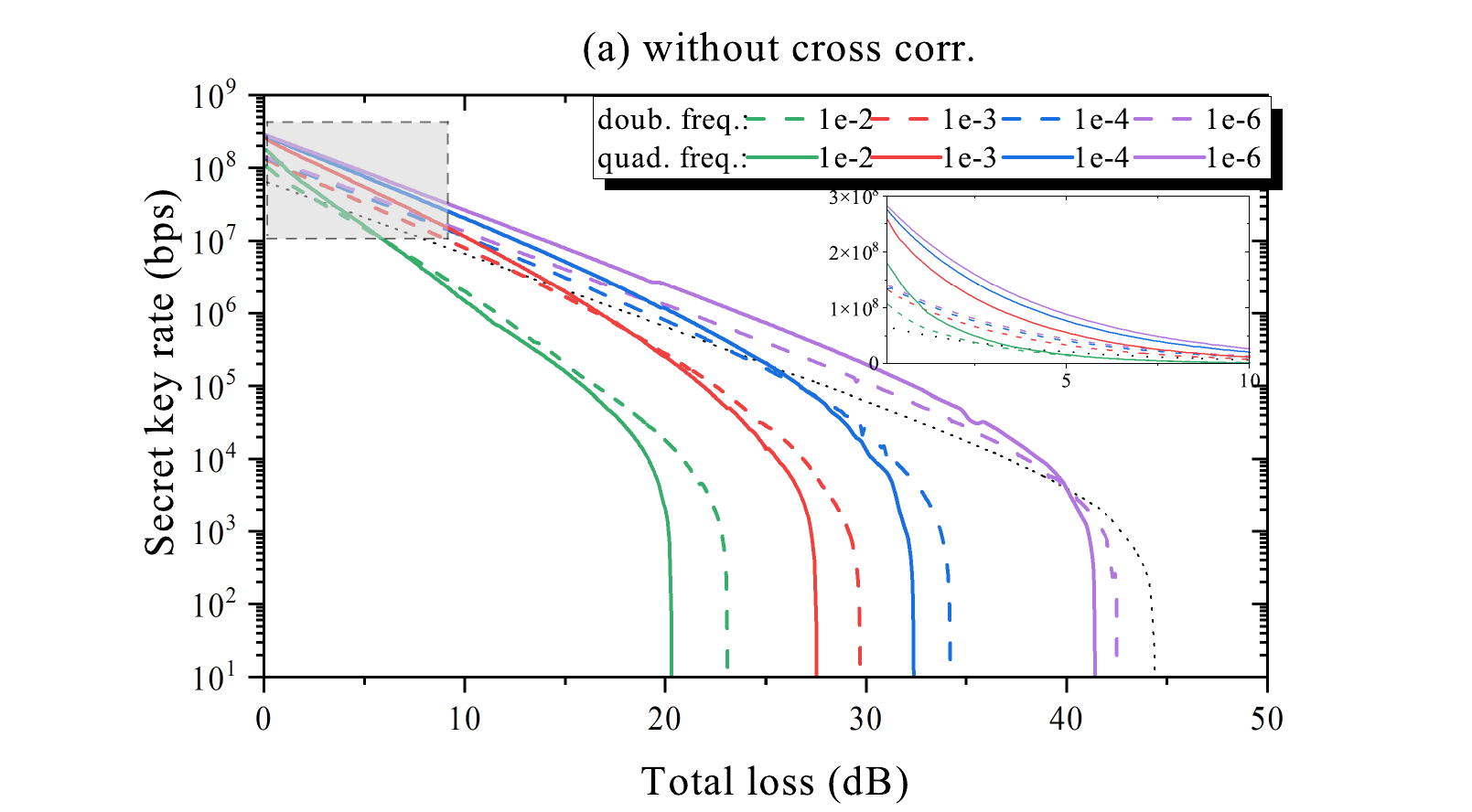}
    \includegraphics[width=0.5\textwidth]{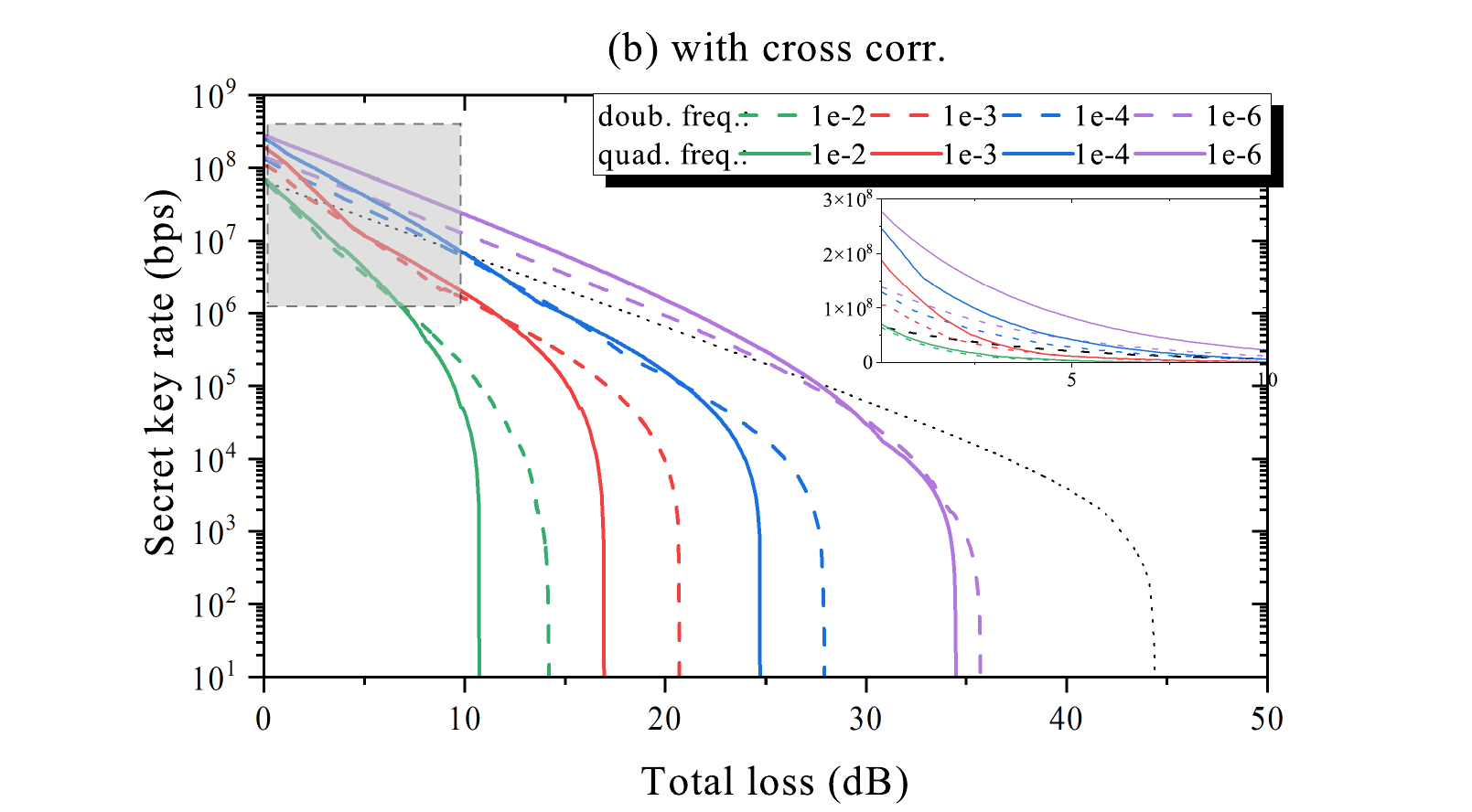}
    \caption{    \label{fig:simu} 
              Secret-key rate (SKR) of the refined decoy-state QKD protocol with an overclocked transmitter. (a) and (b) show the simulation results of a system without and with cross correlations. The black dotted line denotes the SKR of an ideal system, while the colored lines represent the SKR of the double-frequency (dashed lines) and quadruple-frequency (solid lines) systems for different values of the $\varepsilon$ coefficients.        
            } 
\end{figure}

\begin{figure*}[htbp]
	\includegraphics[width=12cm]{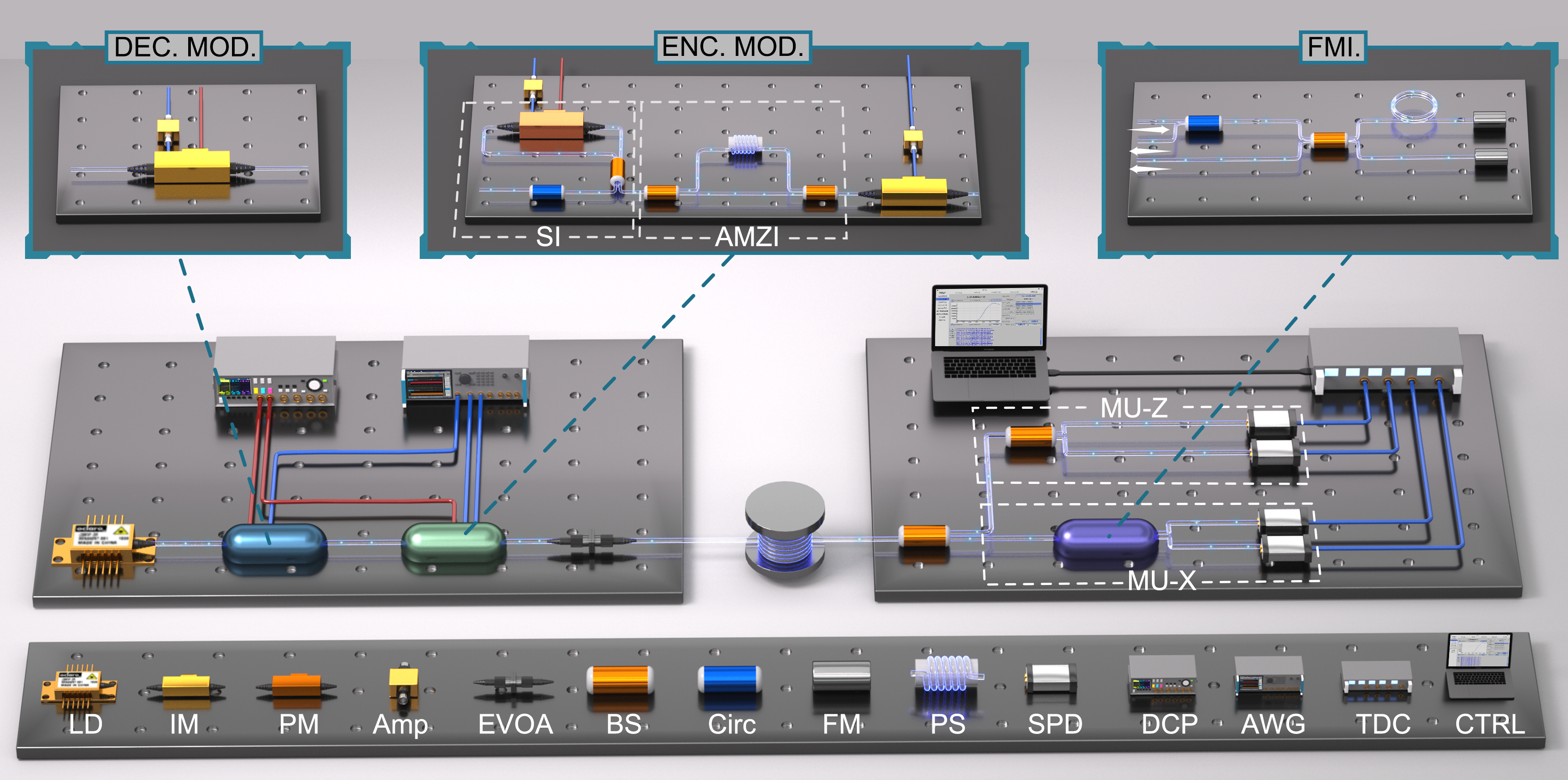 }
	\caption{\label{fig: Exp_setup} Experimental setup for validating our protocol in a overclocked system. LD: laser diode, IM: intensity modulator, OS: optical switch, PM: phase modulator, RFPA: radio frequency power amplifier, EVOA: electronic variable optical attenuator, BS: beam splitter, Circ: circulator, FM: Faraday Mirror, PS: phase shifter, SPD: single-photon detector, DCP: programable DC power source, AWG: arbitrary waveform generator, TDC: time-digital converter, CTRL: controller,
    SI: Sagnac interferometer, AMZI: asymmetric Mach-Zehnder interferometer, FMI: Faraday-Michelson interferometer,
    DEC. MOD.: decoy module, ENC. MOD. encoding module.}
\end{figure*}

\section{Experimental Demonstration of the Qverclocked QKD} To experimentally validate our protocol in the presence of pulse and cross correlations, we employ overclocked devices to build a time-bin-encoding QKD setup operating at 1 GHz (the conservative frequency is 250 MHz). The experiment comprises three parts. The first part involves building the overclocked modulation system and forcing it to operate at the desired 1 GHz repetition rate. In the second part, we accurately characterize the correlations with our `pattern microscope'. Finally, in the third part we suppress the correlations to an ultra-low level by using our `double suppressing' method. We refer the reader to the Methods section for further information.

The experimental setup for the QKD system is illustrated in Fig. \ref{fig: Exp_setup}. On the source side, a gain-switched laser diode (LD, WT-LD100) generates PRWCPs with a pulse width of 50 ps and separated in intervals of 1 ns. 
The pulses are fed to the modulation system, which contains the decoy-state module and the bit/basis encoding module. 
The decoy-state module consists of a high-speed intensity modulator (IM) accompanied with its driving circuit. 
The IM is a commercial $\text{LiNbO}_3$-based integrated Mach-Zehnder interferometer (MZI) that operates at 1 GHz to probabilistically generate the three different intensities. 

In the encoding module, the pulses are first directed to a Sagnac interferometer (SI) \cite{roberts2018patterning,fan2022robust,lu2022unbalanced}, which comprises a customized beamsplitter (BS) with $15:85$ splitting ratio, a phase modulator (PM) placed off-center, and polarization-maintaining fibers for connecting the BS and PM.
The output intensity ratio of the SI at its constructive and destructive interference points is 2:1, and both working points have been proven to be essentially insensitive to electronic disturbance \cite{roberts2018patterning,lu2021intensity}.
The SI operates at its constructive and destructive interference points for the $Z$ and $X$ bases, respectively, serving as a low-correlation IM to balance the intensity of the two bases. 
Following the SI, an AMZI with 500 ps path difference splits each pulses in two---early and late---bins. A phase shifter (PS) is inserted in the long arm of the AMZI to compensate the reference-frame drift \cite{laing2010reference,wang2015phase,lu2020efficient}. After the AMZI, an optical switch (OS)---which is also a commercial $\text{LiNbO}_3$ based integrated MZI---is used to selectively block the pulses based on Alice's encoding: if the bit 0 (1) in the $Z$ basis is selected, the OS blocks the late (early) bin of the signal, whereas both pulses pass through if the encoding $+$ is selected. Note that this encoding process halves the intensity of the $Z$-basis states, this being the reason for employing the preceding SI. After modulation, the pulses are attenuated to the single-photon level and transmitted through the channel. 

At the receiver, a 50:50 BS passively distributes~\cite{boaron2018secure,lu2023experimental,hu2023proof} the incoming pulses into two different measurement units (MUs).
In MU-Z, two homemade SPDs~\cite{he2017sine} are gated \cite{he2017sine,ribordy1998performance,yoshizawa2004gated} at 1 GHz, one being activated during the early bin (bit 0) and the other during the late bin (bit 1). In MU-X, a Faraday-Michelson interferometer (FMI) with the same path difference as the transmitter's AMZI interferes the early and late bins, and two homemade SPDs gated at 1 GHz---to filter out dark counts and inter-round noise---record the $X$-basis bits. The bit 0 (1) corresponds to constructive (destructive) interference. The output signals of the four SPDs are sent to a time-digital converter (TDC) to generate the raw key. 

The modulation system in our experiment has been proven to be bandwidth-limited at 1 GHz frequency in~\cite{kang2022patterning,lu2023intensity}. The correlation length has been measured as $\xi=3$ when operated at 1 GHz \cite{kang2022patterning}. This implies that when implementing conventional protocols like BB84, the conservative modulation frequency is 250 MHz (or less).
In the experiment, we characterize the imperfections with our `deviation microscope' and subsequently strongly mitigate them with our `double suppressing' method. Afterward, we characterize the residual imperfections and use this information to properly set values for the parameters $\varepsilon$ required in the security analysis. Specifically, in the Supplementary Material we define certain classes of parameters $\varepsilon_\Delta$, $\varepsilon_r$, $\hat{\varepsilon}_r$ and $\bar{\varepsilon}_r$ ($\varepsilon_a$, $\hat{\varepsilon}_a$ and $\bar{\varepsilon}_a$) that constrain the impact of previous bit/basis encoding (intensity) settings on the current transmitted state (for formal definitions, see the Supplementary Material).
These $\varepsilon$ parameters are initially computed in a fine-grained manner---\textit{i.e.}, we compute each $\varepsilon$ for all possible sequences of settings---by performing tomography on the intensity of the emitted time bins~\cite{huang2022dependency,huang2025realistic,lu2023intensity}. Note that, in a time-bin scheme, this method allows characterization of not only the actual intensity of each signal but also the actual bit/basis encoding. To simplify subsequent calculations, we conservatively select the worst-case scenario among these fine-grained parameters, an approach that introduces no significant performance degradation.

\begin{figure}[htbp]
	\includegraphics[width=8cm]{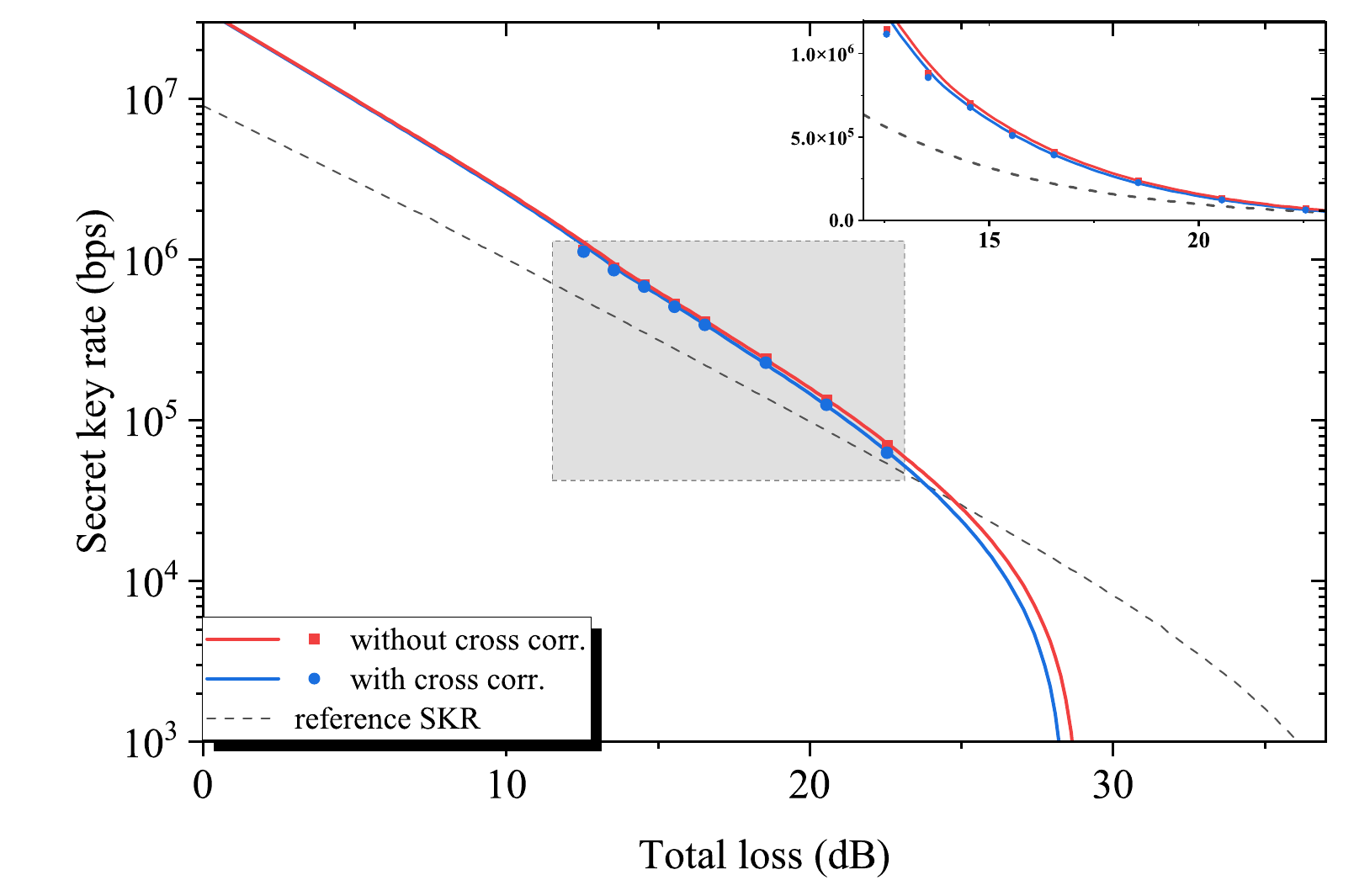 }
	\caption{\label{fig: exp skrs} Experimental and theoretical simulated SKRs. The total loss includes all losses in the quantum channel and detection side. The dashed line corresponds to the ideal BB84 protocol (which matches the performance of the loss-tolerant scheme) operating at 250 MHz repetition rate. 
    The solid blue (red) line corresponds to the overclocked system operating at 1 GHz with (without) cross correlations, with the blue circles (red squares) representing the corresponding experimental results. The subfigure at the upper-right corner is the enlarged view of the gray area, whose linear scale axis indicates that our experimental results achieve a double SKR when working at the overclocked frequency. }
\end{figure}

We operate the setup for several channel losses by employing a 5 km fiber spool and an electronic variable optical attenuator (EVOA). The overall detection efficiency---which comprises all losses at the detection side---is $\sim7\%$ ($\sim 11.55$ dB of overall loss). The raw data is processed under two different scenarios: one in which all correlations are accounted for, and another in which cross-correlations are ignored.  As shown in~\cref{fig: exp skrs}, the experimental results are consistent with the simulations. We obtain a 1.1 Mbps SKR at 5 km, which doubles that of the simulated BB84 (or loss-tolerant \cite{tamaki2014loss,pereira2019quantum}) protocol operated at the safe 250 MHz clock rate. At 11 dB channel loss ($\sim55$ km) we obtain a 69.3 kbps SKR, which is still 1.3 times higher than provided by the simulated ideal system. Indeed, the simulations suggest that our overclocked system would maintain its superiority at intercity distances ($\sim 100$ km) by using SNSPDs~\cite{hao2024compact} at the receiver side.

\section{ CONCLUSION}

In summary, we have proposed a protocol and several techniques to realize the modulation overclock, which is one of the fundamental challenges of QKD systems. On the one hand, the performance of a QKD setup directly depends on its repetition rate, on the other hand, the system must be operated at a limited rate to avoid errors and information leakage due to a correlated and misaligned modulation. 
The protocol presented in this work considers all these potential security loopholes and provides a method to reduce their magnitude to ultra low levels and obtain a much tighter parameter estimation in bandwidth-limited scenarios. 
The simulation results indicate that our protocol allows an overclocked system to achieve a secret-key rate that is several times higher than that of a system operated at the original frequency at metropolitan or intercity distance. 
We have also experimentally demonstrated the protocol by establishing a overclocked system and obtaining a result in agreement with the theory, which confirms the ability to overcome the modulation bandwidth limitation. This study provides a solution to avoid the trade-off between performance and cost in QKD and opens a path towards practical QKD applications.

\begin{acknowledgments}
This work was supported in part by the National Natural Science Foundation of China under Grant 62301524, Grant 62425507, Grant 62271463, Grant 62105318, Grant 61961136004, and Grant 62171424, in part by the Fundamental Research Funds for the Central Universities, and in part by China Postdoctoral
Science Foundation under Grant 2022M723064, in part by the Natural Science Foundation of Anhui under Grant 2308085QF216, and in part by the Innovation Program for Quantum Science and Technology under Grant 2021ZD0300700. M. C. and A. N. acknowledge support from the Galician Regional Government (consolidation of research units: atlanTTic), the Spanish Ministry of Economy and Competitiveness (MINECO), the Fondo Europeo de Desarrollo Regional (FEDER) through the grant No. PID2024-162270OB-I00, MICIN with funding from the European Union NextGenerationEU (PRTR-C17.I1) and the Galician Regional Government with own funding through the “Planes Complementarios de I+D+I con las Comunidades Autonomas” in Quantum Communication, the “Hub Nacional de Excelencia en Comunicaciones Cuanticas” funded by the Spanish Ministry for Digital Transformation and the Public Service and the European Union NextGenerationEU, the European Union’s Horizon Europe Framework Programme under the project “Quantum Security Networks Partnership” (QSNP, grant agreement No 101114043) and the European Union via the European Health and Digital Executive Agency (HADEA) under the Project QuTechSpace (grant 101135225).
\end{acknowledgments}


\appendix


\section{Modulation} To drive Alice's modulators, we employ a 5 GS/s-sampling-rate arbitrary waveform generator (AWG, Tektronix AWG5208) accompanied by a radio frequency power amplifier (RFPA). This electronic system has been proven to be bandwidth-limited at 1 GHz in previous works \cite{kang2022patterning,lu2023intensity}, leading to a correlation range $\xi=3$. 

\begin{figure}[htbp]
	\includegraphics[width=7cm]{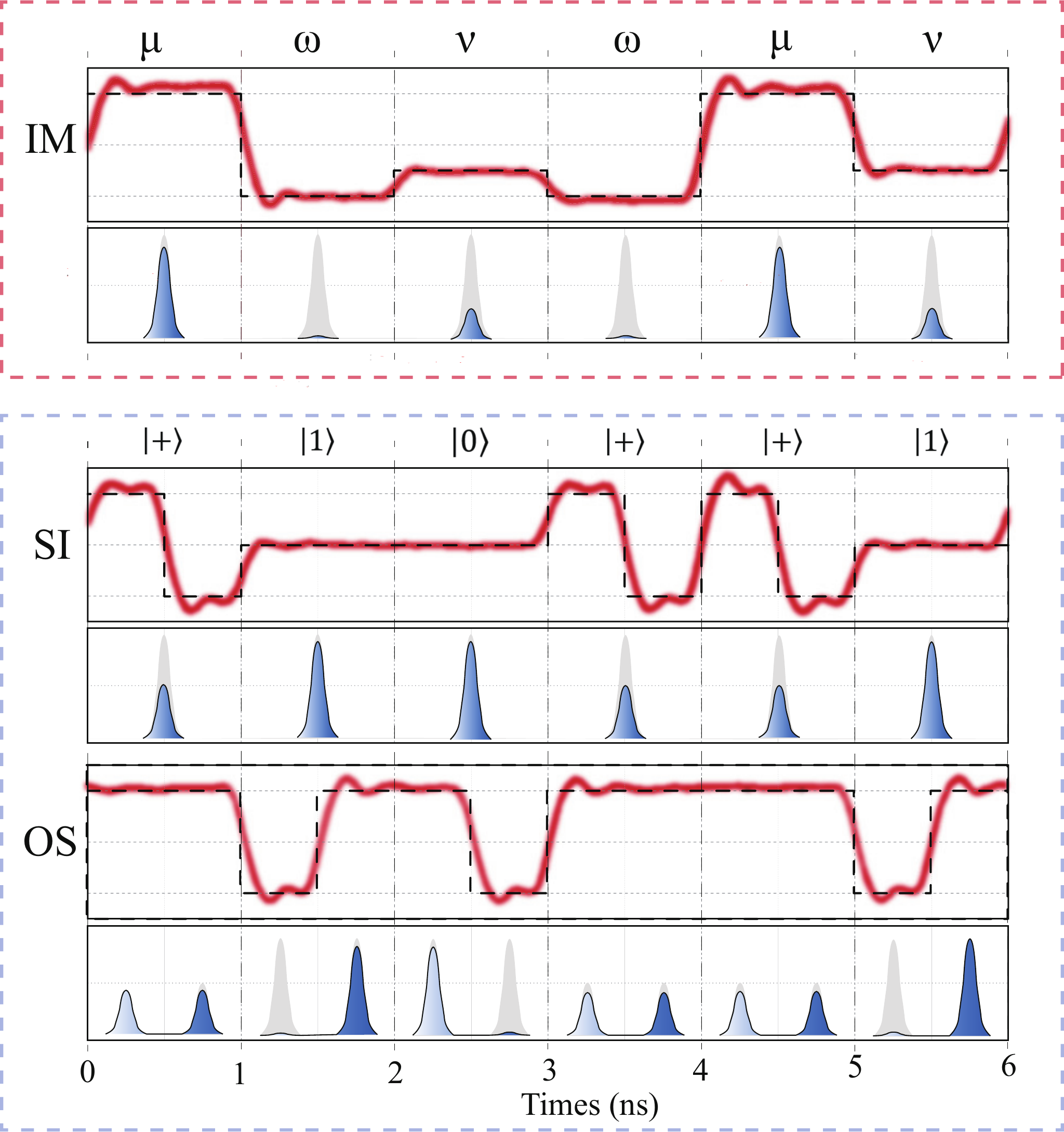}
	\caption{\label{fig: modulation signals} RF signals for the IM, the PM of the SI, and the OS. We consider the intensity sequence $\nu,\mu,\omega,\nu,\omega,\mu $ and the encoding sequence $1, +, +, 0, 1, +$ as an example. Red solid lines and black dashed lines represent the real and ideal RF signals, respectively. The blue (gray) pulses represent the generated coherent pulses after (before) the modulation.}
\end{figure}

As illustrated in the red dashed-line box in Fig. \ref{fig: modulation signals}, the RF signal used for the decoy-state modulation is a square waveform with a duration of 1 ns. The different amplitudes of the modulation signal are fine-tuned by our calibration algorithm according to the full sequence $a_{k-3}^k$. The RF signals used for the bit/basis encoding are illustrated in the blue dashed-line box. In the SI, the clockwise and counterclockwise pulses pass through the PM sequentially, being modulated by positive and negative voltages, respectively. The PM is driven by a calibrated RF signal~\cite{yoshino2018quantum,roberts2018patterning,lu2021intensity} whose amplitude is fine-tuned according to the encoding bases selected in the current and three preceding rounds---which is essentially determined by the settings $r_{k-3}^k$. The amplitude of the $Z$ ($X$) basis corresponds to a relative phase $0$ ($\pi$) between the clockwise and the counterclockwise pulses inside the modulator.  

In contrast to the previous IM , the OS is operated at 2 GHz to independently modulate the early and late time bins in each round.  
The amplitude of its driving RF signal depends on the current configuration of the OS---\texttt{on} or \texttt{off}---and is fine-tuned according to its previous six configurations. 
That is, when the current pulse belongs to a late bin, the amplitude is fine-tuned according to the current early bin and the previous two and a half encoding settings; when the current pulse belongs to an early bin, the amplitude is fine-tuned according to the previous three encoding settings. In particular, the bit/basis encoding settings $0$, $1$, and $+$ correspond to the OS configurations \texttt{on-off}, \texttt{off-on}, and \texttt{on-on}, respectively (see~\cref{fig: modulation signals}). 

\section{Deviation microscope} A precise characterization of the correlations is crucial for fine-tuning the RF signals that are fed to the modulators. In our time-bin scheme, correlations affecting both the bit/basis encoding and the decoy intensity of a signal can be examined by measuring the intensities of its two time bins. However, detecting intensity deviations in weak pulses poses a significant challenge. To overcome this obstacle, we devised a technique we term `deviation microscope'. The key idea is to focus on the most sensitive points of the response curve of the modulators, where intensity deviations can be measured with much higher precision. The technique can be divided in two steps: characterizing the `sensitivity curve' of the modulator, and measuring the output intensity at a sensitive point. Precisely, we define the `sensitivity' as
%
\begin{equation}
    \label{equ:signal-to-deviation ratio}
    \begin{aligned}
       R_{\rm ds}(\beta) = \frac{\abs{I'(\beta)}}{I(\beta) + I_{\rm n}},
   \end{aligned}
\end{equation}
where $\beta$ denotes the working point of the modulator; $I(\beta)$ denotes the normalized (\textit{i.e.}, $\max_{\beta}{I(\beta)}=1$) output of the modulator; $I'(\beta) = \mathrm{d}I(\beta)\big/\mathrm{d} \beta$ is the derivative of $I(\beta)$; and $I_{\rm n}$ denotes the detection noise. Note that~\cref{equ:signal-to-deviation ratio} quantifies the instantaneous rate of change of the output intensity relative to its current value, rather than the absolute rate of change~\cite{yoshino2018quantum,roberts2018patterning,lu2021intensity}.

\begin{figure}
	\includegraphics[width=8cm]{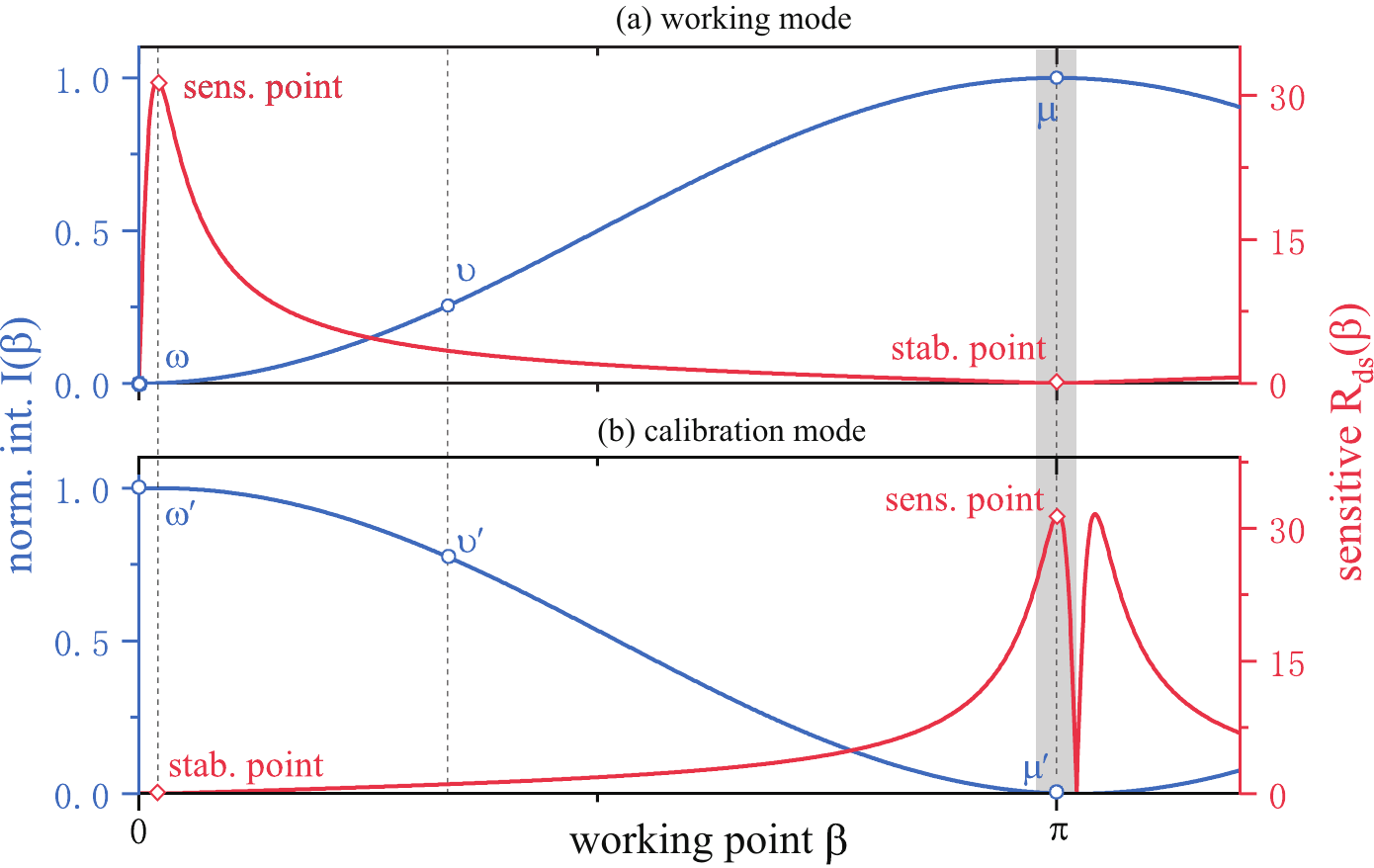}
	\caption{\label{fig: response curve} Response (blue) and sensitivity (red) curves of a commercial MZI-based IM or OS (see~\cref{equ:response curve,equ:signal-to-deviation ratio}). As an example, we consider that Alice measures the deviations of the signal intensity $\mu$ with the correlation microscope. In the working mode, a pre-decided signal corresponding to $\beta$ is loaded to the RF port of the IM (OS) to prepare the desired intensity. In the calibration mode, an additional bias $\beta_{\rm b}$ is loaded to the bias port of the IM (OS) to move the target intensity to a sensitive point. Once the deviations are characterized, the bias voltage is removed, thus returning to the working mode.}
\end{figure}

Let us consider a commercial MZI-based IM or OS as an example. The response curve of this type of devices follows a sinusoidal form~\cite{ye2023induced,lu2023hacking}:
\begin{equation}
    \label{equ:response curve}
    \begin{aligned}
       I(\beta) = \alpha_{\rm in} \left[  \cos \left( \beta + \beta_{\rm b} \right) + 1 + I_{\rm b} \right] / 2    ,
   \end{aligned}
\end{equation}
where $\alpha_{\rm in}$ is the input intensity, $\beta_{\rm b}$ denotes the bias of the modulator \cite{ye2023induced,lu2023hacking}, and $I_{\rm b}$ represents the unavoidable background intensity. As shown in~\cref{fig: response curve}, the working point with highest sensitivity is very close to the vacuum intensity $\omega$. Unfortunately, measuring the intensity near the vacuum point is challenging in practice because the output signal is typically overwhelmed by noise, substantially reducing the signal-to-noise ratio (SNR). Indeed, this is the primary reason why the correlations of the vacuum intensity have been neglected in previous studies. 

\begin{figure}[htbp]
	\includegraphics[width=8cm]{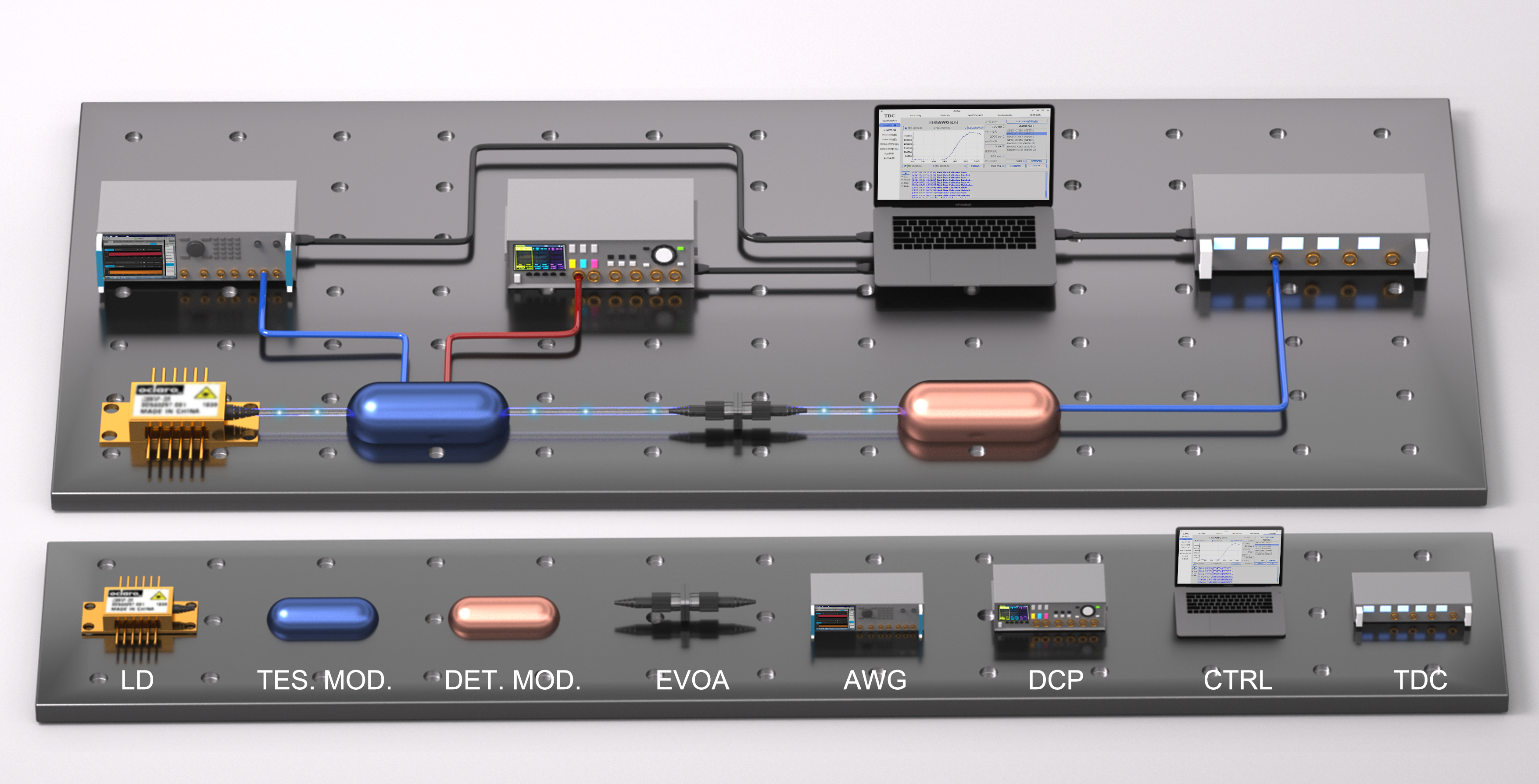}
	\caption{\label{fig: calib setup} Schematic of our correlation measurent system. LD: laser diode, TES. MOD.: tested module, DET. MOD. detection module, EVOA: electronic variable optical attenuator, DCP: programmable DC power source, AWG: arbitrary waveform generator, TDC: time-digital converter, CTRL: controller.}
\end{figure}

To address this challenge, we built a setup to test the decoy module and encoding modules independently. The setup is depicted in Fig. \ref{fig: calib setup}, where the blue capsule represents the currently tested module, and the brown capsule represents a detection module that can be adapted to the tested module.
The calibration process proceeds as follows.   
Each round, a RF signal $V_a$ chosen at random from the pre-decided set $\left\{ V_\mu, V_\nu, V_\omega \right\}$ is fed into the IM to determine its working point $\beta$.
To measure the correlations of the target intensity, say $\mu$, the IM is biased to make $\beta(V_\mu)$ correspond to a highly sensitive point according to~\cref{equ:signal-to-deviation ratio}.  In the detection module, an SPD is employed due to its proven superiority in detecting weak signals~\cite{eraerds2010photon,kirmani2014first,shin2016photon}. The SPD is gated such that it is only active when the intensity setting $\mu$ is selected, which filters out dark counts and afterpulses. Moreover, this selective gating effectively filters out unwanted responses from non-target intensities, preventing detector count saturation. As a result, it allows the user to reduce the attenuation of the EVOA (see Fig. \ref{fig: calib setup}), thereby enhancing the SNR and enabling the observation of correlations even at weak intensities. The fine-grained detection statistics at the sensitive point are then classified according to the previous settings and subsequently used to calibrate the RF signals. Finally, the original bias voltage is recovered to obtain the correlation-suppressed signals.


\begin{figure}[htbp]
	\includegraphics[width=8cm]{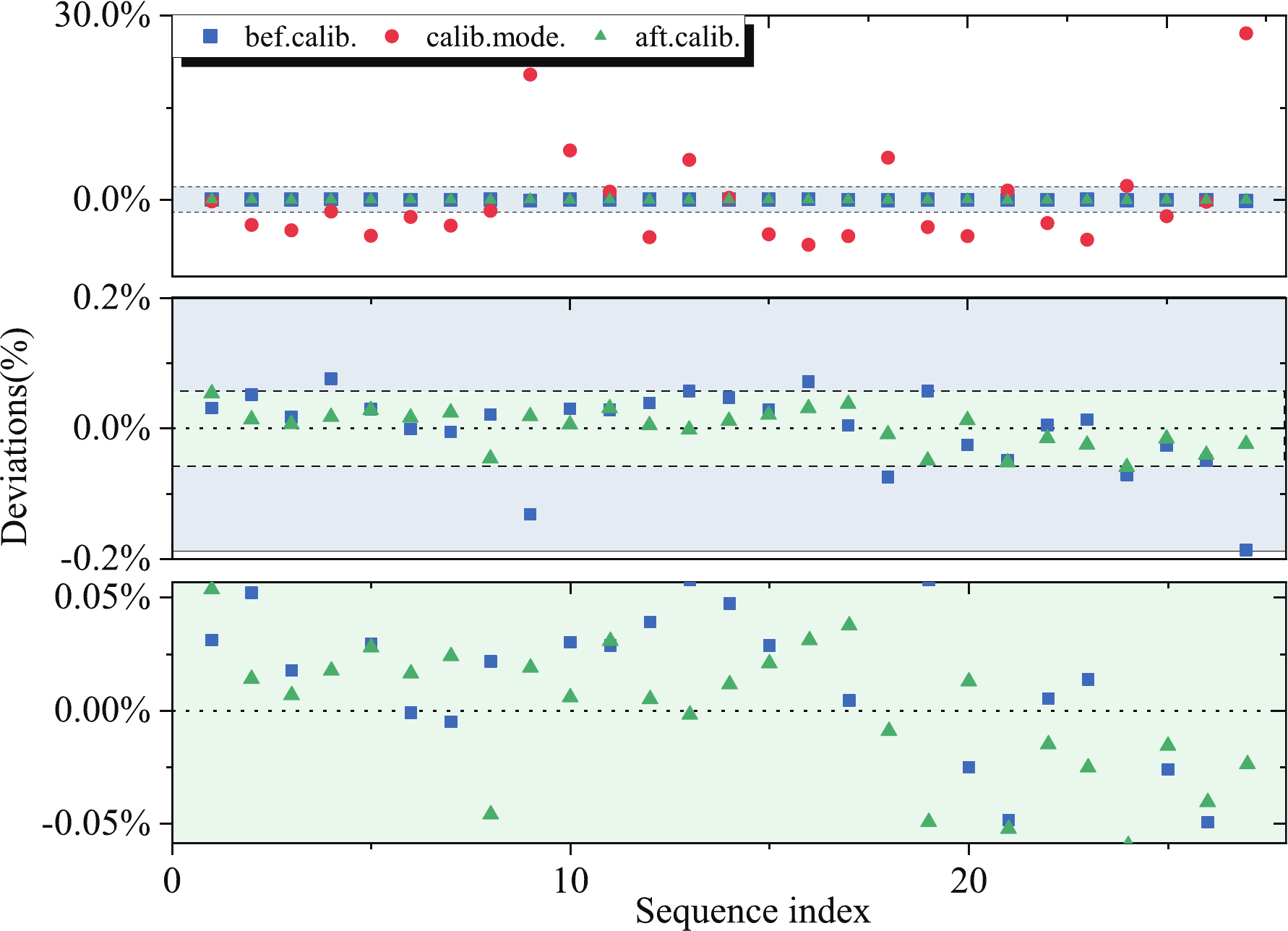 }
	\caption{\label{fig: exp correlation mag glass} Experimental results with the deviation microscope and double suppressing. The blue squares and green triangles represent the deviations before and after compensating the RF signal, respectively, while the red circles denote the deviations observed with the deviation microscope. Each integer in the x axis represents a different pattern $a_{k-3}^{k-1}$. Specifically, the pattern index is computed as $9{a_{k-1}} + 3{a_{k-2}} + {a_{k-3}}$, with $\omega =0$, $\nu =1$, and $\mu = 2$. The blue and green areas contain the squares and triangles, respectively. The top sub-figure shows the distribution of different deviations, the middle sub-figure is the enlarged view of the blue area, and the bottom sub-figure is the enlarged view of the green area.}
\end{figure}

\section{Double suppressing:} Previous studies have demonstrated that optical stable points significantly mitigate the intensity deviations caused by correlations to a level as low as $0.2\%$ \cite{roberts2018patterning,lu2021intensity}. Moreover, an electronic compensating algorithm have been demonstrated to suppress such deviations to a level of $1\%$~\cite{kang2022patterning,lu2023intensity}. In this work, we address a critical technical challenge, which is the fact that the intensity deviations at the optical stable points are too small to be reliably estimated for the compensating algorithms. By overcoming this limitation, we achieve the double suppression that combines the optical solution and the electronic compensation algorithm.

Specifically, the target modulator is first biased to leverage the deviation microscope, allowing the intensity deviations to be observed at sensitive points. Subsequently, the compensating algorithm is executed to suppress the deviations by adjusting the RF signal. After the algorithm is completed, the additional bias is removed, ensuring that the output remains at the stable point while the distortions are compensated. This method can further reduce the deviations at the stable point by an order of magnitude. We employ it to suppress the deviations at the IM and OS, achieving an exceptionally low level of deviations. For the remaining active components of the transmitter, we simply employ either the deviation microscope or the compensating algorithm~\cite{lu2023intensity}.

In the experiment, we employ Bob's SPDs to measure the deviations. Since our homemade SPDs can only operate at 1 GHz, we reduce the pulse rate of the laser source to 200 MHz while keeping the modulators operating at their original frequency. This means that each `pulse round' is followed by four `empty rounds'. To characterize the deviations for the target intensity, say $\mu$, we load $V_{\mu}$ into the IM in the `pulse rounds' and select random intensities in the `empty rounds'. As the correlation range is three, this procedure allows to observe all relevant patterns. Then, we classify the measurement statistics according to the previous three selections $a_{k-3}^{k-1}$ and compute the yield $D_{\mu,a_{k-3}^{k-1}}$---\textit{i.e.}, the probability to observe a detection given that Alice selects the settings $\mu$ and the previous three settings $a_{k-3}^{k-1}$---for each group. We use here the letter $D$ to differentiate the yield in the correlation measurements from that of the QKD experiments (for which we use the letter $Y$). As illustrated in~\cref{fig: exp correlation mag glass}, prior to compensation, the deviations for $\mu$ are all below $2\%$, which is an acceptable range. This is possible because $\mu$ is generated at a stable point of the IM. To further suppress the deviations, we bias the IM to measure the yields at a sensitive point. There, we observe significant deviations on the order of $10\%$, with a maximum deviation reaching $27.1\%$. By compensating the RF signal at this point and removing the bias voltage, we finally obtain deviations of $\sim0.02\%$, with maximum at $0.052\%$. This demonstrates a level of performance that is state-of-the-art when compared to similar works.


\bibliography{citations}

\end{document}